\documentstyle[12pt]{article}

\catcode`\@=11

\begin{document}

\newlength{\lno} \lno1.5cm \newlength{\len} \len=\textwidth%
\addtolength{\len}{-\lno}

\baselineskip7mm \renewcommand{\thefootnote}{\fnsymbol{footnote}} \newpage

\begin{titlepage}
\vspace{2.0cm}
\begin{center}
{\Large\bf Quantum  Lax Pair  From Yang-Baxter Equations}\\
\vspace{1.5cm}
{\large A. Lima-Santos }\footnote{e-mail: dals@power.ufscar.br} \\
\vspace{1.0cm}
{\large \em Universidade Federal de S\~ao Carlos, Departamento de F\'{\i}sica \\
Caixa Postal 676, CEP 13569-905~~S\~ao Carlos, Brasil}\\
\end{center}
\vspace{3.0cm}

\begin{abstract}
We show explicitly how to construct the quantum Lax pair for systems
with open boundary conditions. We demonstrate the method by applying it
to the  Heisenberg XXZ model with the general integrable boundary terms.
\end{abstract}
\vfill
\end{titlepage}

\baselineskip6mm

\newpage{}

Ever since the introduction of the quantum Lax pair into statistical
mechanics by McCoy and Wu \cite{Wu}\ and of the notion of a one-parameter
family of commuting transfer matrices by Baxter \cite{Baxter1}, a great deal
of effort has been expended in the search for spin chains with Hamiltonian $%
{\cal H}$ and two-dimensional statistical models with transfer matrix $\tau $
which have the integrability property.

The quantum inverse scattering method places the theory of completely
integrable quantum system and solvable statistical models in a unified
framework \cite{Faddeev}. The basis to apply it to a completely integrable
system is to associate an operator version of an auxiliary linear problem 
\cite{Wadati}: 
\begin{equation}
\Psi _{n+1}=L_{n}\Psi _{n}\ \quad ,\quad \ \stackrel{.}{\Psi }_{n}=A_{n}\Psi
_{n},  \label{eq1}
\end{equation}
where $L_{n}$ and $A_{n}$ are matrix operators depending on the spectral
parameter $u$, and a dot signifies a time derivative. The consistency
condition for equations (\ref{eq1}) with $\stackrel{.}{u}=0$ yields the Lax
pair equation: 
\begin{equation}
\stackrel{.}{L}_{n}=A_{n+1}L_{n}-L_{n}A_{n}.  \label{eq2}
\end{equation}
All the solved integrable models appear to imply that \ a model is
completely integrable if we can find a Lax pair $\left\{ L_{n},A_{n}\right\} 
$ \ such that the Lax equation (\ref{eq2}) is equivalent to the equation of
motion of the model \cite{Wadati}.

Starting with a local Lax operator $L_{n}$ (a matrix acting in the auxiliary
space $V$ with elements acting in the quantum space $h_{n}$, at site $n$),
for most quantum integrable system the product direct of two Lax operators $%
L_{n}$, with different spectral parameters satisfy a similarity relation 
\begin{equation}
{\cal R}(u-v)L_{n}(u)\otimes L_{n}(v)=L_{n}(v)\otimes L_{n}(u){\cal R}(u-v)
\label{eq3}
\end{equation}
with ${\cal R}$ a $c$-number matrix acting in $V\otimes V$. The above
relation is referred to as the local Yang-Baxter relation and can be
represented graphically by figure $1$.

\begin{center}
\setlength{\unitlength}{5947sp}\begingroup\makeatletter\ifx\SetFigFont%
\undefined\gdef\SetFigFont#1#2#3#4#5{\reset@font\fontsize{#1}{#2pt} %
\fontfamily{#3}\fontseries{#4}\fontshape{#5} \selectfont}\fi\endgroup%
\begin{picture}(3024,1050)(1189,-1723)
\thinlines
\put(1201,-1411){\line( 2, 1){1200}}
\put(1201,-961){\line( 2,-1){1200}}
\put(2101,-661){\line( 0,-1){1050}}
\put(3001,-811){\line( 2,-1){1200}}
\put(3001,-1561){\line( 2, 1){1200}}
\put(3301,-661){\line( 0,-1){1050}}
\put(1445,-1331){\makebox(0,0)[lb]{\smash{\SetFigFont{6}{7.2}{\rmdefault}{\mddefault}{\updefault}$\cal{R}$$(u$-$v)$}}}
\put(2051,-1036){\makebox(0,0)[lb]{\smash{\SetFigFont{6}{7.2}{\rmdefault}{\mddefault}{\updefault}$L_{n}(v)$}}}
\put(2051,-1411){\makebox(0,0)[lb]{\smash{\SetFigFont{6}{7.2}{\rmdefault}{\mddefault}{\updefault}$L_{n}(u)$}}}
\put(2976,-1036){\makebox(0,0)[lb]{\smash{\SetFigFont{6}{7.2}{\rmdefault}{\mddefault}{\updefault}$L_{n}(u)$}}}
\put(2976,-1400){\makebox(0,0)[lb]{\smash{\SetFigFont{6}{7.2}{\rmdefault}{\mddefault}{\updefault}$L_{n}(v)$}}}
\put(3576,-1361){\makebox(0,0)[lb]{\smash{\SetFigFont{6}{7.2}{\rmdefault}{\mddefault}{\updefault}$\cal{R}$$(u$-$v)$}}}
\end{picture}

figure $1$
\end{center}

In terms of the operators $L_{n}$, the monodromy matrix for a chain with
length $N$ is expressed as 
\begin{equation}
T(u)=L_{N}(u)L_{N-1}(u)\cdots L_{2}(u)L_{1}(u)  \label{eq4}
\end{equation}
and if $L_{n}^{\prime }$s with different $n$ commute, we further have 
\begin{equation}
{\cal R}(u-v)T(u)\otimes T(v)=T(v)\otimes T(u){\cal R}(u-v)  \label{eq5}
\end{equation}
called global Yang-Baxter relation.

We assume the auxiliary space $V$ and the quantum space $h$ are the same and
regard the elements of $L$ and ${\cal R}$ as those of the $R$-matrix (${\cal %
R}={\cal P}R$ with ${\cal P}$ the permutation operator which interchanges
spaces of $V_{n}$ with $V_{n-1}$ ), in terms of which the local Yang-Baxter
relation (\ref{eq3}) has the form 
\begin{equation}
R_{12}(u-v)R_{13}(u)R_{23}(v)=R_{23}(v)R_{13}(u)R_{12}(u-v),  \label{eq5a}
\end{equation}
where $R_{ij}$ is a matrix in $V\times V\times V$, which acts non-trivially
in the spaces $V_{i}$ and $V_{j}$ only.

In two-dimensional statistical mechanics, the elements of $R$ may be
considered as vertices of a vertex model. Baxter first noticed the
importance of the relation (\ref{eq5a}) in that context and regarded it as
the solvability condition of the vertex model \cite{Baxter1}: we introduce
the transfer matrix $\tau (u)$ as a trace of the monodromy matrix $\tau (u)=%
{\rm tr}T(u)$. The relation (\ref{eq5}) indicates that there exists a family
of commuting transfer matrices and that $u$-expansion of $\tau (u)$ gives a
set of conserved quantities which are involutive.

The integrability conditions for a system with open boundary condition are
formulated in order that both the Yang-Baxter equation and the boundary
Yang-Baxter equations (or reflection equations) are satisfied \cite{Sklyanin}%
. To a quantum system on a finite interval with independent boundary
conditions at each end, we have to introduce reflection matrices $K^{\mp
}(u) $ to describe such boundary conditions. For a $PT$-invariant $R$%
-matrix, the fundamental reflection-factorization relations obeyed by $%
K^{-}(u)$ and $K^{+}(u)$\ are \cite{Mezin}: 
\begin{equation}
R_{12}(u-v)K_{1}^{-}(u)R_{21}(u+v)K_{2}^{-}(v)=K_{2}^{-}(v)R_{12}(u+v)K_{1}^{-}(u)R_{21}(u-v)
\label{eq6}
\end{equation}
which is represented graphically by figure $2$

\begin{center}
{}\setlength{\unitlength}{4947sp}\begingroup\makeatletter\ifx\SetFigFont%
\undefined\gdef\SetFigFont#1#2#3#4#5{\reset@font\fontsize{#1}{#2pt} %
\fontfamily{#3}\fontseries{#4}\fontshape{#5} \selectfont}\fi\endgroup%
\begin{picture}(2475,1674)(1801,-2023)
\thinlines
\put(2401,-361){\line( 0,-1){1650}}
\put(2101,-1861){\line( 2, 5){300}}
\put(2401,-1111){\line(-1, 2){300}}
\put(1951,-1711){\line( 2, 1){450}}
\put(2401,-1486){\line(-2, 1){450}}
\put(1951,-1261){\line( 0, 1){  0}}
\put(1951,-1261){\line( 0, 1){  0}}
\put(1951,-1261){\line( 0, 1){  0}}
\put(4201,-361){\line( 0,-1){1650}}
\put(3901,-1861){\line( 2, 5){300}}
\put(4201,-1111){\line(-1, 2){300}}
\put(3751,-1036){\line( 2, 1){450}}
\put(4201,-811){\line(-2, 1){450}}
\put(3751,-586){\line( 0, 1){  0}}
\put(3751,-586){\line( 0, 1){  0}}
\put(3751,-586){\line( 0, 1){  0}}
\put(1826,-1670){\makebox(0,0)[lb]{\smash{\SetFigFont{6}{7.2}{\rmdefault}{\mddefault}{\updefault}$u$}}}
\put(2026,-1861){\makebox(0,0)[lb]{\smash{\SetFigFont{6}{7.2}{\rmdefault}{\mddefault}{\updefault}$v$}}}
\put(3826,-1861){\makebox(0,0)[lb]{\smash{\SetFigFont{6}{7.2}{\rmdefault}{\mddefault}{\updefault}$v$}}}
\put(3626,-976){\makebox(0,0)[lb]{\smash{\SetFigFont{6}{7.2}{\rmdefault}{\mddefault}{\updefault}$u$}}}
\put(2326,-1486){\makebox(0,0)[lb]{\smash{\SetFigFont{6}{7.2}{\rmdefault}{\mddefault}{\updefault}$K^{\scriptstyle{-}}_{\scriptstyle{1}}(u)$}}}
\put(2326,-1111){\makebox(0,0)[lb]{\smash{\SetFigFont{6}{7.2}{\rmdefault}{\mddefault}{\updefault}$K^{\scriptstyle{-}}_{\scriptstyle{2}}(v)$}}}
\put(4126,-811){\makebox(0,0)[lb]{\smash{\SetFigFont{6}{7.2}{\rmdefault}{\mddefault}{\updefault}$K^{\scriptstyle{-}}_{\scriptstyle{1}}(u)$}}}
\put(4126,-1111){\makebox(0,0)[lb]{\smash{\SetFigFont{6}{7.2}{\rmdefault}{\mddefault}{\updefault}$K^{\scriptstyle{-}}_{\scriptstyle{2}}(v)$}}}
\put(1801,-1231){\makebox(0,0)[lb]{\smash{\SetFigFont{6}{7.2}{\rmdefault}{\mddefault}{\updefault}-$u$}}}
\put(1951,-491){\makebox(0,0)[lb]{\smash{\SetFigFont{6}{7.2}{\rmdefault}{\mddefault}{\updefault}-$v$}}}
\put(3601,-566){\makebox(0,0)[lb]{\smash{\SetFigFont{6}{7.2}{\rmdefault}{\mddefault}{\updefault}-$u$}}}
\put(3751,-491){\makebox(0,0)[lb]{\smash{\SetFigFont{6}{7.2}{\rmdefault}{\mddefault}{\updefault}-$v$}}}
\end{picture}

figure $2$
\end{center}

For the left boundary we have a similar equation: 
\begin{eqnarray}
&&R_{12}(-u+v)K_{1}^{+}(u)^{t_{1}}M_{1}^{-1}R_{21}(-u-v-2\rho
)M_{1}K_{2}^{+}(v)^{t_{2}}  \nonumber \\
&=&K_{2}^{+}(v)^{t_{2}}M_{1}R_{12}(-u-v-2\rho
)M_{1}^{-1}K_{1}^{+}(u)^{t_{1}}R_{21}(-u+v)  \label{eq7}
\end{eqnarray}
Here $t_{i}$ denotes the transposition in the $i^{{\it th}}$ space; $%
K_{1}^{\pm }=K^{\pm }\otimes 1,K_{2}^{\pm }=1\otimes K^{\pm }$ , etc. $\rho $
is a crossing parameter and $M$ is a crossing matrix, both being specific to
a given $R$-matrix. These relations correspond to the constraint of
factorized scattering in the presence of a wall.

There is an automorphism \cite{Sklyanin, Mezin} between the $K^{-}$ and $%
K^{+}$ namely, given a solution $K^{-}(u)$, then $K^{-}(-u-\rho )^{t}M$ is a
solution $K^{+}(u)$.

Given an $R$ -matrix and $K$-matrices satisfying (\ref{eq6}) and (\ref{eq7}%
), the corresponding open chain transfer matrix $t(u)$ is given by 
\begin{equation}
t(u)={\rm tr}\left( K^{+}(u)T(u)K^{-}(u)T^{-1}(-u)\right)  \label{eq8}
\end{equation}
which constitutes a one-parameter commutative family.

Hamiltonians \ with boundary terms are obtained from the first derivative of
the transfer matrix (\ref{eq8}) \cite{Sklyanin}: 
\begin{equation}
{\cal H}=\sum_{n=2}^{N}H_{n,n-1}+\frac{1}{2}\frac{K_{1}^{-\prime }(0)}{%
K_{1}^{-}(0)}+\frac{{\rm tr}_{a}[K_{a}^{+}(0)H_{Na}]}{{\rm tr}K^{+}(0)}
\label{eq9}
\end{equation}
where the local sum is given by 
\begin{equation}
\sum_{n=2}^{N}H_{n,n-1}=\alpha \left. \frac{{\rm d}}{{\rm d}u}\ln \tau
(u)\right| _{u=0}+{\rm const\ }I  \label{eq9a}
\end{equation}
The index $a$ denotes the auxiliary space and $\alpha $ is a constant.

The corresponding Lax pair formulation may be obtained by direct
consideration of the equations of motion which can be write in the Lax form
considering the following operator version of an auxiliary linear problem 
\begin{equation}
\Psi _{n+1}=L_{n}\Psi _{n}\qquad n=1,2,...,N\quad ,\quad \stackrel{.}{\Psi }%
_{n}=A_{n}\Psi _{n}\qquad n=2,3,...,N  \label{eq11}
\end{equation}
and 
\begin{equation}
\stackrel{.}{\Psi }_{1}=Q_{1}\Psi _{1}\quad ,\quad \stackrel{.}{\Psi }%
_{N+1}=Q_{N}\Psi _{N+1}.  \label{eq12}
\end{equation}
Note that $Q_{1}(u)$ and $Q_{N}(u)$ are responsible for the boundary terms
in the equations of motion of the system. The consistency conditions for
these equations are the following Lax equations: 
\begin{equation}
\stackrel{.}{L}_{n}=A_{n+1}L_{n}-L_{n}A_{n}\quad {\rm for\quad }n=2,...,N-1.
\label{eq13}
\end{equation}
and 
\begin{equation}
\stackrel{.}{L}_{1}=A_{2}L_{1}-L_{1}Q_{1}\quad {\rm and}\quad \stackrel{.}{L}%
_{N}=Q_{N}L_{N}-L_{N}A_{N}  \label{eq14}
\end{equation}

These equations specify the evolution only of $L_{n}$. \ It means that $%
A_{n} $ , $Q_{1}$ and $Q_{N}$ \ have to be determined in terms of $L_{n}$
and $H_{n,n-1}$.

In this note we will show how to find the Lax pair from the boundary
Yang-Baxter equations. In this way we are including the general integrable
boundary terms in a previous work of Zhang \cite{Zhang} where the periodic
case was considered. We also apply the method \ to the open {\small XXZ}
model and compare the result with previous calculations.

To do this we consider a application of the local Yang-Baxter relation,
where the vertices are being commuted horizontally rather than vertically: 
\begin{equation}
{\cal R}_{n,n-1}(\epsilon )L_{n}(u+\epsilon
)L_{n-1}(u)=L_{n}(u)L_{n-1}(u+\epsilon ){\cal R}_{n,n-1}(\epsilon ).
\label{eq21}
\end{equation}
Here the spectral parameters of the local $L$-operators differ only by an
infinitesimal amount $\epsilon $. This relation can be represented
graphically by figure $3$

\begin{center}
\setlength{\unitlength}{2947sp}\begingroup\makeatletter\ifx\SetFigFont%
\undefined\gdef\SetFigFont#1#2#3#4#5{\reset@font\fontsize{#1}{#2pt} %
\fontfamily{#3}\fontseries{#4}\fontshape{#5} \selectfont}\fi\endgroup%
\begin{picture}(6924,2129)(2089,-3073)
\thinlines
\put(2101,-2461){\line( 1, 0){2700}}
\put(3901,-961){\line(-3,-4){1548}}
\put(4501,-3061){\line(-3, 4){1548}}
\put(8701,-961){\line(-3,-4){1548}}
\put(8108,-3056){\line(-3, 4){1548}}
\put(6301,-1561){\line( 1, 0){2700}}
\put(6751,-1411){\makebox(0,0)[lb]{\smash{\SetFigFont{6}{7.2}{\rmdefault}{\mddefault}{\updefault}$L_{n}(u)$}}}
\put(8326,-1411){\makebox(0,0)[lb]{\smash{\SetFigFont{6}{7.2}{\rmdefault}{\mddefault}{\updefault}$L_{n-1}(u$+$\epsilon)$}}}
\put(2001,-2986){\makebox(0,0)[lb]{\smash{\SetFigFont{6}{7.2}{\rmdefault}{\mddefault}{\updefault}$n$}}}
\put(3850,-2986){\makebox(0,0)[lb]{\smash{\SetFigFont{6}{7.2}{\rmdefault}{\mddefault}{\updefault}$n$-$1$}}}
\put(6826,-2986){\makebox(0,0)[lb]{\smash{\SetFigFont{6}{7.2}{\rmdefault}{\mddefault}{\updefault}$n$}}}
\put(8001,-2986){\makebox(0,0)[lb]{\smash{\SetFigFont{6}{7.2}{\rmdefault}{\mddefault}{\updefault}$n$-$1$}}}
\put(2551,-2661){\makebox(0,0)[lb]{\smash{\SetFigFont{6}{7.2}{\rmdefault}{\mddefault}{\updefault}$L_{n}(u$+$\epsilon)$}}}
\put(4151,-2661){\makebox(0,0)[lb]{\smash{\SetFigFont{6}{7.2}{\rmdefault}{\mddefault}{\updefault}$L_{n-1}(u)$}}}
\put(3351,-1661){\makebox(0,0)[lb]{\smash{\SetFigFont{6}{7.2}{\rmdefault}{\mddefault}{\updefault}$\cal{R}$$_{n,n-1}(\epsilon)$}}}
\put(6450,-2461){\makebox(0,0)[lb]{\smash{\SetFigFont{6}{7.2}{\rmdefault}{\mddefault}{\updefault}$\cal{R}$$_{n,n-1}(\epsilon)$}}}
\end{picture}

figure $3$
\end{center}

This is a form of the star-triangle equation used in \cite{Zhang, Thacker,
Helen}. This formulation follows also from the non-local quantum Lax pair 
\cite{BP, ALS}.

Using the normalization ${\cal R}(0)=1$ we have the following $\epsilon $%
-expansions 
\begin{eqnarray}
{\cal R}(\epsilon ) &\sim &1+\epsilon (\alpha ^{-1}H+\beta I)+o(\epsilon
^{2}),  \label{eq22a} \\
L_{n}(u+\epsilon ) &\sim &L_{n}(u)+\epsilon L_{n}^{\prime }(u)+o(\epsilon
^{2}),  \label{eq22b}
\end{eqnarray}
where a prime stands for $u$-derivative, with $\alpha $ and $\beta $ some
constants and $H_{n,n-1}$ of the local sum in (\ref{eq9a}) is the $H$ in (%
\ref{eq22a}) acting on the quantum spaces at sites $n$ and $n-1$.

Substituting (\ref{eq22a}) and (\ref{eq22b}) into (\ref{eq21}) we get as the
first non-trivial consequence the following commutation relation 
\begin{eqnarray}
\alpha ^{-1}[H_{n,n-1},L_{n}(u)L_{n-1}(u)] &=&L_{n}(u)L_{n-1}^{\prime
}(u)-L_{n}^{\prime }(u)L_{n-1}(u)  \nonumber \\
n &=&2,3,...,N  \label{eq23}
\end{eqnarray}
We multiply out the above equation to arrive at 
\begin{eqnarray}
\lbrack H_{n,n-1},L_{n}] &=&\alpha (L_{n}L_{n-1}^{\prime
}L_{n-1}^{-1}-L_{n}^{\prime })-L_{n}[H_{n,n-1},L_{n}]L_{n-1}^{-1}
\label{eq24} \\
\lbrack H_{n+1,n},L_{n}] &=&\alpha (L_{n}^{\prime
}-L_{n+1}^{-1}L_{n+1}^{\prime }L_{n})-L_{n+1}^{-1}[H_{n+1,n},L_{n+1}]L_{n}
\label{eq25}
\end{eqnarray}
for $n=2,3,...,N-1$.

In order to include the boundary terms into (\ref{eq24}) and (\ref{eq25}) we
also shall consider a reflection equation where the vertices are being
commuted horizontally: 
\begin{equation}
\left[ L_{1}(u)L_{0}(u+2\epsilon )K_{1}^{-}(\epsilon )\right]
K_{2}^{-}(u+\epsilon )=\left[ K_{1}^{-}(\epsilon )L_{1}(u+2\epsilon
)L_{0}(u)\right] K_{2}^{-}(u+\epsilon )  \label{eq26}
\end{equation}
This equation can be represented graphically by figure $4$

\begin{center}
\setlength{\unitlength}{2947sp}\begingroup\makeatletter\ifx\SetFigFont%
\undefined\gdef\SetFigFont#1#2#3#4#5{\reset@font\fontsize{#1}{#2pt} %
\fontfamily{#3}\fontseries{#4}\fontshape{#5} \selectfont}\fi\endgroup%
\begin{picture}(8325,1935)(1351,-3586)
\thinlines
\put(1501,-3361){\line( 1, 0){3000}}
\put(1501,-2461){\line( 5,-3){1500}}
\put(3001,-3361){\line( 5, 3){1500}}
\put(1501,-1861){\line( 1,-2){750}}
\put(2251,-3361){\line( 1, 2){750}}
\put(6601,-3361){\line( 1, 0){3000}}
\put(8131,-1846){\line( 1,-2){750}}
\put(8881,-3346){\line( 1, 2){750}}
\put(6601,-2461){\line( 5,-3){1500}}
\put(8101,-3361){\line( 5, 3){1500}}
\put(1301,-1711){\makebox(0,0)[lb]{\smash{\SetFigFont{6}{7.2}{\rmdefault}{\mddefault}{\updefault}$u$+$\epsilon$}}}
\put(2801,-1786){\makebox(0,0)[lb]{\smash{\SetFigFont{6}{7.2}{\rmdefault}{\mddefault}{\updefault}-$u$-$\epsilon$}}}
\put(1151,-2461){\makebox(0,0)[lb]{\smash{\SetFigFont{6}{7.2}{\rmdefault}{\mddefault}{\updefault}$\epsilon$}}}
\put(4376,-2386){\makebox(0,0)[lb]{\smash{\SetFigFont{6}{7.2}{\rmdefault}{\mddefault}{\updefault}-$\epsilon$}}}
\put(1776,-3586){\makebox(0,0)[lb]{\smash{\SetFigFont{6}{7.2}{\rmdefault}{\mddefault}{\updefault}$K^{\scriptstyle{-}}_{2}(u$+$\epsilon)$}}}
\put(2801,-3586){\makebox(0,0)[lb]{\smash{\SetFigFont{6}{7.2}{\rmdefault}{\mddefault}{\updefault}$K^{\scriptstyle{-}}_{1}(\epsilon)$}}}
\put(6326,-2386){\makebox(0,0)[lb]{\smash{\SetFigFont{6}{7.2}{\rmdefault}{\mddefault}{\updefault}$\epsilon$}}}
\put(7901,-1711){\makebox(0,0)[lb]{\smash{\SetFigFont{6}{7.2}{\rmdefault}{\mddefault}{\updefault}$u$+$\epsilon$}}}
\put(9401,-1711){\makebox(0,0)[lb]{\smash{\SetFigFont{6}{7.2}{\rmdefault}{\mddefault}{\updefault}-$u$-$\epsilon$}}}
\put(9476,-2461){\makebox(0,0)[lb]{\smash{\SetFigFont{6}{7.2}{\rmdefault}{\mddefault}{\updefault}-$\epsilon$}}}
\put(7626,-3586){\makebox(0,0)[lb]{\smash{\SetFigFont{6}{7.2}{\rmdefault}{\mddefault}{\updefault}$K^{\scriptstyle{-}}_{1}(\epsilon)$}}}
\put(8701,-3586){\makebox(0,0)[lb]{\smash{\SetFigFont{6}{7.2}{\rmdefault}{\mddefault}{\updefault}$K^{\scriptstyle{-}}_{2}(u$+$\epsilon)$}}}
\put(1751,-2611){\makebox(0,0)[lb]{\smash{\SetFigFont{6}{7.2}{\rmdefault}{\mddefault}{\updefault}$L_{1}(u)$}}}
\put(2406,-2986){\makebox(0,0)[lb]{\smash{\SetFigFont{6}{7.2}{\rmdefault}{\mddefault}{\updefault}$L_{0}(u$+$2\epsilon)$}}}
\put(7551,-2961){\makebox(0,0)[lb]{\smash{\SetFigFont{6}{7.2}{\rmdefault}{\mddefault}{\updefault}$L_{1}(u$+$2\epsilon)$}}}
\put(9076,-2836){\makebox(0,0)[lb]{\smash{\SetFigFont{6}{7.2}{\rmdefault}{\mddefault}{\updefault}$L_{0}(u)$}}}
\end{picture}

figure $4$
\end{center}

Using the the normalization $K^{-}(0)=1$ and 
\begin{equation}
K^{-}(\epsilon )\sim 1+\epsilon K^{-^{\prime }}(0)+o(\epsilon ^{2}),
\label{eq27}
\end{equation}
we get from (\ref{eq26}) a new commutation relation 
\begin{equation}
\lbrack \frac{1}{2}K_{1}^{-\prime
}(0),L_{1}(u)L_{0}(u)]=L_{1}(u)L_{0}^{\prime }(u)-L_{1}^{\prime }(u)L_{0}(u)
\label{eq28}
\end{equation}
We multiply out the above equation and arrive at 
\begin{equation}
\lbrack \frac{1}{2}K_{1}^{-\prime }(0),L_{1}]=(L_{1}L_{0}^{\prime
}L_{0}^{-1}-L_{1}^{\prime })-L_{1}[\frac{1}{2}K_{1}^{-\prime
}(0),L_{1}]L_{0}^{-1}  \label{eq29}
\end{equation}
Note that in these equations we have extended the chain in order to include
the site $n=0$ at the right boundary and made use of the non-singular
property of the $L$-operators.

Now we recall the equation (\ref{eq24}) to see that the identification 
\begin{equation}
\alpha ^{-1}H_{1,0}=\frac{1}{2}K_{1}^{-\prime }(0)\quad  \label{eq210}
\end{equation}
allows its continuation for $n=1$.

By similar considerations one can derive the left boundary relations. Now
the equation (\ref{eq25}) is continued for $n=N$ with the identification 
\begin{equation}
\alpha ^{-1}H_{N+1,N}=\frac{{\rm tr}\left[ K_{a}^{+}(0)H_{Na}\right] }{{\rm %
tr}K^{+}(0)}  \label{eq211}
\end{equation}

These results tell us that we can keep all considerations made by Zhang for
the periodic case\cite{Zhang}. The main difference consist in remove the
term $H_{N,N+1}=H_{N,1}$ from the periodic Hamiltonian and adding two
boundary terms $H_{1,0}$ and $H_{N+1,N}$ determined by the matrices $K^{\pm
}(u)$.

The equation of motion for $L_{n}$ is the Heisenberg equation 
\begin{equation}
\stackrel{.}{L}_{n}={\rm i}[{\cal H},L_{n}]={\rm i}[H_{n+1,n},L_{n}]+{\rm i}%
[H_{n,n-1},L_{n}],  \label{eq212}
\end{equation}
here we have set $\hbar =1$. The second equality is because of the locality
of the ${\cal H}$ and $L_{n}$. Combining (\ref{eq24}),(\ref{eq25}) and (\ref
{eq212}), we have 
\begin{equation}
\stackrel{.}{L}_{n}\!=\!-{\rm i}L_{n+1}^{-1}\!\left\{ \!\alpha
L_{n+1}^{\prime }\!+\![H_{n+1,n},L_{n+1}]\!\right\} \!L_{n}\!+\!{\rm i}%
L_{n}\!\left\{ \!\alpha L_{n-1}^{\prime }\!-\![H_{n,n-1},L_{n-1}]\!\right\}
\!L_{n-1}^{-1}  \label{eq213}
\end{equation}
By comparing with the Lax equation (\ref{eq13}), we can read off the second
Lax operator 
\begin{equation}
A_{n}=-{\rm i}L_{n}^{-1}\!\left\{ \alpha L_{n}^{\prime }\!+\left[
H_{n,n-1},L_{n}\right] \!\!\right\} \!  \label{eq214}
\end{equation}
or 
\begin{equation}
A_{n}=-{\rm i}\left\{ \!\alpha L_{n-1}^{\prime
}\!-\![H_{n,n-1},L_{n-1}]\!\right\} \!L_{n-1}^{-1}  \label{eq215}
\end{equation}
The compatibility of (\ref{eq214}) and (\ref{eq215}) is guaranteed by the
commutation relation (\ref{eq24}).

Therefore, the second Lax operator for completely integrable open chains has
the following form: In the bulk it is the same for the corresponding
periodic chain 
\begin{eqnarray}
A_{n} &=&{\rm i}H_{n,n-1}-{\rm i}L_{n}^{-1}H_{n,n-1}L_{n}-{\rm i}\alpha
L_{n}^{-1}\!L_{n}^{\prime }\!  \nonumber \\
n &=&2,3,...,N,  \label{eq216}
\end{eqnarray}
at the right boundary it is given by 
\begin{equation}
Q_{1}={\rm i}\alpha \left\{ \frac{1}{2}K_{1}^{-\prime }(0)-\frac{1}{2}%
L_{1}^{-1}K_{1}^{-\prime }(0)L_{1}-L_{1}^{-1}\!L_{1}^{\prime }\right\}
\label{eq217}
\end{equation}
and at the left boundary it is read off from the equation (\ref{eq215}): 
\begin{equation}
Q_{N}={\rm i}H_{N+1,N}-{\rm i}L_{N}\ H_{N+1,N}\ L_{1}^{-1}-{\rm i}\alpha
L_{N}^{\prime }\!\ L_{N}^{-1},  \label{eq218}
\end{equation}
where $H_{N+1,N}$ is given by (\ref{eq211}).

Finally, we shall apply this method to a concrete model, the one-dimensional
Heisenberg {\small XXZ} open chain with Hamiltonian \cite{Sklyanin, deVega}: 
\begin{eqnarray}
{\cal H} &=&-\sum_{k=2}^{N}\left( \sigma _{k}^{+}\sigma _{k-1}^{-}+\sigma
_{k}^{-}\sigma _{k-1}^{+}+\frac{1}{2}\cos 2\eta \ \sigma _{k}^{z}\sigma
_{k-1}^{z}\right)  \nonumber \\
&&+\sin 2\eta \left( A_{-}\sigma _{1}^{z}+B_{-}\sigma _{1}^{+}+C_{-}\sigma
_{1}^{-}+A_{+}\sigma _{N}^{z}+B_{+}\sigma _{N}^{+}+C_{+}\sigma
_{N}^{-}\right)  \label{eq31}
\end{eqnarray}
where 
\begin{equation}
A_{\mp }=\frac{1}{2}\cot (\xi _{\mp })\ ,\ B_{\mp }=\frac{b_{\mp }}{\sin \xi
_{\mp }}\ ,\ C_{\mp }=\frac{c_{\mp }}{\sin \xi _{\mp }}  \label{eq32}
\end{equation}
Here $\sigma ^{x},\sigma ^{y},\sigma ^{z}$ and $\sigma ^{\pm }=\frac{1}{2}%
(\sigma ^{x}\pm {\rm i}\sigma ^{y})$ are the usual Pauli spin-$\frac{1}{2}$
operators. $\eta $ is a parameter associated with the anisotropy, $b_{\mp }$%
, $c_{\mp }$ and $\xi _{\mp }$ are some constants describing the boundary
effects.

It is not difficult to verify that the equations of motion derived from the
Hamiltonian (\ref{eq31}) may be cast in the Lax form (\ref{eq13}) and (\ref
{eq14}).

The first Lax operator $L_{n}$ is identified with the $R$-matrix of the $6$%
-vertex model \cite{KS} 
\begin{eqnarray}
L_{n} &=&R_{na}=\left( 
\begin{array}{ccc}
w_{4}+w_{3}\sigma _{n}^{z} &  & 2w_{1}\sigma _{n}^{-} \\ 
&  &  \\ 
2w_{1}\sigma _{n}^{+} &  & w_{4}-w_{3}\sigma _{n}^{z}
\end{array}
\right)  \nonumber \\
&=&w_{4}+w_{3}\sigma ^{z}\sigma _{n}^{z}+2w_{1}(\sigma ^{-}\sigma
_{n}^{+}+\sigma ^{+}\sigma _{n}^{-})  \label{eq33}
\end{eqnarray}
where the elements are parametrized by trigonometric functions of $u$%
\begin{equation}
w_{4}+w_{3}=\sin (u+2\eta ),\quad w_{4}-w_{3}=\sin u,\quad 2w_{1}=\sin 2\eta
\label{eq34}
\end{equation}
Using the $\epsilon $-expansion (\ref{eq22a}) we have 
\begin{equation}
\alpha =-\sin 2\eta \quad {\rm and}\quad \beta =-\frac{1}{2}\cot 2\eta
\label{eq35}
\end{equation}
The inverse and the first derivative of $L_{n}$ are well-defined: 
\begin{eqnarray}
L_{n}^{-1} &=&v_{4}+v_{3}\sigma ^{z}\sigma _{n}^{z}+2v_{1}\left( \sigma
^{-}\sigma _{n}^{+}+\sigma ^{+}\sigma _{n}^{-}\right)  \nonumber \\
L_{n}^{\prime } &=&w_{4}^{\prime }+w_{3}^{\prime }\sigma ^{z}\sigma _{n}^{z}
\label{eq36}
\end{eqnarray}
where 
\begin{eqnarray}
v_{4}+v_{3} &=&\frac{1}{\Delta }\sin (u-2\eta ),\quad v_{4}-v_{3}=\frac{1}{%
\Delta }\sin u,\quad 2v_{1}=-\frac{1}{\Delta }\sin 2\eta  \nonumber \\
\Delta &=&\sin (u-2\eta )\sin (u+2\eta )  \label{eq37}
\end{eqnarray}
Now, by direct substitution of these data in (\ref{eq216}) one can easily
find the second Lax operator $A_{n}$ \ for $n=2,3,...,N$ : 
\begin{eqnarray}
A_{n} &=&{\rm iconst.}I+{\rm i}d(u)\sigma _{n}^{z}\sigma _{n-1}^{z}+{\rm i}%
f(u)\left( \sigma _{n}^{-}\sigma _{n-1}^{+}+\sigma _{n}^{+}\sigma
_{n-1}^{-}\right)  \nonumber \\
&&+{\rm i}\sigma ^{z}\left[ g(u)\left( \sigma _{n}^{-}\sigma
_{n-1}^{+}-\sigma _{n}^{+}\sigma _{n-1}^{-}\right) -d(u)\left( \sigma
_{n}^{z}+\sigma _{n-1}^{z}\right) \right]  \nonumber \\
&&-{\rm i}\sigma ^{+}\left[ p(u)\left( \sigma _{n}^{-}\sigma
_{n-1}^{z}-\sigma _{n}^{z}\sigma _{n-1}^{-}\right) +q(u)\left( \sigma
_{n}^{-}+\sigma _{n-1}^{-}\right) \right]  \nonumber \\
&&-{\rm i}\sigma ^{-}\left[ p(u)\left( \sigma _{n}^{z}\sigma
_{n-1}^{+}-\sigma _{n}^{+}\sigma _{n-1}^{z}\right) +q(u)\left( \sigma
_{n}^{+}+\sigma _{n-1}^{+}\right) \right]  \label{eq38}
\end{eqnarray}
where 
\begin{eqnarray}
d(u) &=&\frac{1}{4\Delta }\sin 4\eta \sin 2\eta ,\quad f(u)=\frac{1}{\Delta }%
\sin \eta \left( \sin \eta \cos 2u+\sin 3u\right) ,  \nonumber \\
g(u) &=&\frac{1}{2\Delta }\sin 2\eta \sin 2u,\quad p(u)=\frac{1}{2\Delta }%
\sin 4\eta \sin u,\quad q(u)=\frac{1}{2\Delta }\sin 4\eta \sin u  \nonumber
\\
{\rm const.} &=&\frac{1}{2\Delta }\sin 2\eta (\sin 2u-\sin 2\eta \cos 2\eta )
\label{eq39}
\end{eqnarray}
This solution is the Lax operator for the periodic {\small XXZ \ }spin chain
which was obtained by Sogo and Wadati \cite{SW} in the trigonometric limit
of the Lax pair operators for the one-dimensional {\small XYZ} Heisenberg
spin chain.

Now we have to compute the corresponding boundary operators. By a direct
calculation with the operators $L_{1}$ and $L_{1}^{-1}$ given by (\ref{eq36}%
) and together with (\ref{eq210}), one gets the following results for each
term of $Q_{1}$: 
\begin{equation}
{\rm i}H_{10}=-{\rm i}\alpha \left( A_{-}\sigma _{1}^{z}+B_{-}\sigma
_{1}^{+}+C_{-}\sigma _{1}^{-}\right)  \label{eq310}
\end{equation}
\begin{eqnarray}
-{\rm i}\alpha L_{1}^{-1}L_{1} &=&-{\rm i}\alpha \left( v_{4}w_{4}^{\prime
}+v_{3}w_{3}^{\prime }\right) -{\rm i}\alpha \left( v_{4}w_{3}^{\prime
}+v_{3}w_{4}^{\prime }\right) \sigma ^{z}\sigma _{1}^{z}  \nonumber \\
&&-2{\rm i}v_{1}\left( w_{4}^{\prime }-w_{3}^{\prime }\right) \left( \sigma
^{-}\sigma _{1}^{+}+\sigma ^{+}\sigma _{1}^{-}\right)  \label{eq311}
\end{eqnarray}
and 
\begin{eqnarray}
&&-{\rm i}L_{1}^{-1}H_{01}L_{1}={\rm i}\alpha
A_{-}(v_{4}w_{4}+v_{3}w_{3}-2v_{1}w_{1})\sigma _{1}^{z}+{\rm i}\alpha
A_{-}(v_{4}w_{3}+v_{3}w_{4}+2v_{1}w_{1})\sigma ^{z}  \nonumber \\
&&+2{\rm i}\alpha A_{-}\left[ (v_{4}-v_{3})w_{1}-v_{1}(w_{4}-w_{3})\right]
(\sigma ^{-}\sigma _{1}^{+}-\sigma ^{+}\sigma _{1}^{-})  \nonumber \\
&&+{\rm i}\alpha (v_{4}w_{4}-v_{3}w_{3})(B_{-}\sigma _{1}^{+}+C_{-}\sigma
_{1}^{-})+{\rm i}\alpha (v_{3}w_{4}-v_{4}w_{3})\sigma ^{z}(B_{-}\sigma
_{1}^{+}-C_{-}\sigma _{1}^{-})  \nonumber \\
&&+{\rm i}\alpha \left[ w_{1}(v_{4}+v_{3})+v_{1}(w_{4}+w_{3})\right]
(B_{-}\sigma ^{+}+C_{-}\sigma ^{-})  \nonumber \\
&&+{\rm i}\alpha \left[ w_{1}(v_{4}+v_{3})-v_{1}(w_{4}+w_{3})\right]
(B_{-}\sigma ^{+}-C_{-}\sigma ^{-})\sigma _{1}^{z}  \label{eq312}
\end{eqnarray}
We thus find the following Lax operator 
\begin{eqnarray}
Q_{1} &=&{\rm iconst.}{\bf 1}+\frac{{\rm i}\sin ^{2}2\eta }{\Delta \sin \xi
_{-}}\times  \nonumber \\
&&  \nonumber \\
&&\!\!\!\!\!\!\!\!\!\!\!\left( 
\begin{array}{ccc}
\begin{array}{c}
-\frac{1}{2}\sin (2\eta +\xi _{-})\sigma _{1}^{z}+\frac{1}{2}\cos \xi
_{-}\sin 2\eta \\ 
+b_{-}t(u)\sigma _{1}^{+}+c_{-}r(u)\sigma _{1}^{-}
\end{array}
&  & 
\begin{array}{c}
\sin (u-\xi _{-})\sigma _{1}^{-} \\ 
+b_{-}\sin 2\eta \cos u-c_{-}\sin u\cos 2\eta \sigma _{1}^{z}
\end{array}
\\ 
&  &  \\ 
\begin{array}{c}
-\sin (u+\xi _{-})\sigma _{1}^{+} \\ 
+b_{-}\sin 2\eta \cos u+c_{-}\sin u\cos 2\eta \sigma _{1}^{z}
\end{array}
&  & 
\begin{array}{c}
-\frac{1}{2}\sin (2\eta -\xi _{-})\sigma _{1}^{z}-\frac{1}{2}\cos \xi
_{-}\sin 2\eta \\ 
+b_{-}r(u)\sigma _{1}^{+}+c_{-}t(u)\sigma _{1}^{-}
\end{array}
\end{array}
\right)  \nonumber \\
&&  \label{eq314}
\end{eqnarray}
where 
\begin{equation}
t(u)=\frac{\sin u\cos (u-\eta )}{\cos \eta }-\sin 2\eta \quad {\rm and}\quad
r(u)=\frac{\sin u\cos (u+\eta )}{\cos \eta }-\sin 2\eta .  \label{eq315}
\end{equation}
The left boundary operator $Q_{N}$ is obtained using the equation (\ref
{eq218}). This results that $\ Q_{N}$ \ is the transposition of $Q_{1}$,
followed by the following substitution: 
\begin{equation}
\xi _{-}\rightarrow \xi _{+},\quad b_{-}\rightarrow c_{+},\quad
c_{-}\rightarrow b_{+}\quad {\rm and}\quad \sigma _{1}\rightarrow \sigma _{N}
\label{eq316}
\end{equation}
These Lax operators are the trigonometric limit of the Lax pair for the open 
{\small XYZ} spin chain given in ref.\cite{Ju}. In particular, the diagonal
case ($b_{\mp }=c_{\mp }=0$), was first derived in \cite{Zhou}.

To summarize: solving the Yang-Baxter equation together with the reflection
equations, we can read off $L_{n}$ , ${\cal R}$ and $K^{\mp }$ operators; if 
$\ {\cal H}$ has the form (\ref{eq9}) with its bulk (\ref{eq1}) is related
to ${\cal R}$ by (\ref{eq22a}), then the corresponding $A_{n}$ operator is
given by (\ref{eq216}). Moreover, for every solution $\ K^{-}$ ($K^{+}$)\ of
the reflection equation (\ref{eq6}) ((\ref{eq7})) with $K^{-}(0)\neq 0$ ($%
{\rm Tr}K^{+}(0)\neq 0$), we can read off the right (left) boundary term of $%
{\cal H}$ which has the form (\ref{eq210})\ ((\ref{eq211})), then the
corresponding $Q_{1}$ ($Q_{N}$) operator is given by (\ref{eq217}) ((\ref
{eq218})).

\vspace{0.5cm}{}

{\bf Acknowledgment:} This work was supported in part by Funda\c{c}\~{a}o de
Amparo \`{a} Pesquisa do Estado de S\~{a}o Paulo--FAPESP--Brasil and by
Conselho Nacional de Desenvol\-{}vimento--CNPq--Brasil.

\end{document}